\documentclass[11pt]{article}
\usepackage{overpic,bbm,bm,ulem}
\usepackage{braket}
\usepackage{hyperref}
\usepackage{amssymb}
\usepackage{amsthm}
\usepackage{amsxtra}
\usepackage{color}
\usepackage{multirow}
\usepackage{epsfig}
\usepackage{diagbox}    
\usepackage{amsmath}
\usepackage{booktabs}
\usepackage[]{caption}
\usepackage[column=O]{cellspace}
\usepackage{multirow}
\usepackage{cite}

\begin{document}

\begin{center}
{\Large Three-zero textures of neutrino mass matrix and leptogenesis \\ in the left-right symmetric model}
\end{center}

\vspace{0.05cm}

\begin{center}
{\bf Ding-Hui Xu, \bf Zhen-hua Zhao\footnote{Corresponding author: zhaozhenhua@lnnu.edu.cn}, \bf Tian-Rui Wang} \\
{ $^1$ Department of Physics, Liaoning Normal University, Dalian 116029, China \\
$^2$ Center for Theoretical and Experimental High Energy Physics, \\ Liaoning Normal University, Dalian 116029, China }
\end{center}

\vspace{0.2cm}

\begin{abstract}
Within the framework of the left-right symmetric model (LRSM) and under the assumption of a diagonal Dirac neutrino mass matrix $M_{\rm D}$,
this paper systematically investigates 20 types of three-zero textures in the Majorana neutrino mass matrix $M_{\rm R}$.
The study reveals that only five three-zero textures of $M_{\rm R}$ satisfy the constraints of the latest NuFit 6.0 global fit results. 
Furthermore, we phenomenologically explore the correlations between Majorana phases $\rho$ and $\sigma$, 
as well as the relationships between the heavy neutrino mass spectrum $M_{I}~(I=1,2,3)$, ratio of Dirac matrix elements $y_{2}$, $y_{3}$ and the scale factor $r$. The results indicate strong correlations among the model parameters. 
In particular, the allowed regions for the Majorana CP phases are significantly restricted and depend on the specific texture of $M_{\rm R}$.
On this basis, leptogenesis originating from heavy right-handed neutrino decays is investigated. 
Numerical results demonstrate that the $M_{\nu3}$ pattern can achieve successful leptogenesis within specific $r$ intervals for both the normal ordering (NO) and the inverted ordering (IO) of the light neutrino masses, while the $M_{\nu4}$ and $M_{\nu5}$ patterns possess viable parameter space for successful leptogenesis only in the NO case.
\end{abstract}

\newpage

\section{Introduction}
The discovery of neutrino flavor oscillations represents a monumental achievement in particle physics over the past two decades~\cite{1,2,3}, demonstrating that neutrinos are massive and that the lepton mixing matrix is non-trivial.
However, the dynamical origin of the light neutrino masses and lepton flavor mixing is elusive for theorists.
Fortunately, the quark sector offers a highly instructive precedent, namely well-motivated texture zeros in the quark mass matrix enable the derivation of simple, testable relations between flavor mixing angles and quark mass ratios~\cite{4,5,6,7,8}, in the lack of a convincing flavor theory.
These scenarios are more popularly known as zero texture models, a comprehensive research of which within three neutrino framework can be found in the review article~\cite{9}.

Currently, the seesaw mechanism stands as the most plausible and prevalent framework for accounting for the tiny masses of neutrinos.
Its most standard implementation is the type-I seesaw scenario~\cite{10,11,12,13,14}, which extends the Standard Model (SM) by introducing three right-handed neutrinos $N_{I}~(I=1,2,3)$.
These right-handed neutrinos couple to the left-handed lepton doublets $l_{\alpha} = (\nu_{\alpha}, \alpha)^{T}$ ($\alpha = e, \mu, \tau$)
and the Higgs doublet $H = (H^{+}, H^{0})^{T}$ via the Yukawa coupling term $(Y_{\nu})_{\alpha I}\,\overline{l_{\alpha}}\,H\,N_{I}$, where $Y_{\nu}$ denotes the neutrino Yukawa coupling matrix. 
Following electroweak symmetry breaking (EWSB) (when the Higgs field acquires a vacuum expectation value (VEV) $v$ = 174~GeV),
they induce the Dirac mass matrix $(M_{\rm D})_{\alpha I} = (Y_{\nu})_{\alpha I}\, v$, where $M_{\rm D}$ is the Dirac mass matrix of neutrinos.
Furthermore, since $N_{I}$ are singlets under the SM gauge group, 
they can naturally accommodate a Majorana mass term $(M_{\rm R})_{IJ}\,\overline{N_I^c}\,N_J$ that is independent of EWSB, where $M_{\rm R}$ is the Majorana mass matrix of the right-handed neutrinos. 
In the seesaw limit where $M_{\rm R}\gg M_{\rm D}$, the heavy degrees of freedom are integrated out, resulting in an effective mass matrix for the light neutrinos: $M_{\nu} \simeq - M_{\rm D}\, M_{\rm R}^{-1}\, M_{\rm D}^{T}$.
In addition, the seesaw mechanisms also include the type-II seesaw~\cite{15,16,17,18} and the type-III seesaw~\cite{19}, which introduce scalar and fermion triplets, respectively.

Beyond neutrino mass generation, the seesaw framework offers a compelling origin for the observed baryon asymmetry of the Universe~\cite{20}, known as the leptogenesis mechanism~\cite{21,22,23,24,25}. The magnitude of this asymmetry is typically quantified as:
\begin{equation}
\label{eq:1}
Y_{\rm B} = \frac{n_{\rm B} - n_{\bar{\rm B}}}{s} \simeq (8.68 \pm 0.06)\times 10^{-11},
\end{equation}
where $n_{\rm B}$ ($n_{\bar{\rm B}}$) represents the baryon (antibaryon) number density and $s$ denotes the entropy density.
In this scenario: CP-violating, out-of-equilibrium decays of the heavy right-handed neutrinos first create a primordial lepton asymmetry $Y_{\rm L} \equiv (n_{\rm L} - n_{\bar{\rm L}})/s$. Subsequently, this lepton number is partially reprocessed into a baryon asymmetry via electroweak sphaleron transitions,
leading to the relation $Y_{\rm B} \simeq c_{1} Y_{\rm L}$ with $c_{1}$ $= -$ 28/79.

Recently, a very popular new physics framework beyond standard model (BSM) is the LRSM~\cite{26,27,28,29,30,31,32,33,34,35,36,37,38,39,40}, 
in which the gauge symmetry of the SM is extended to $ \rm SU(3)_C\times SU(2)_L\times SU(2)_R\times U(1)_{B-L}$, so the right-handed fermions, which are singlets in the SM, can form doublets under the new $\rm SU(2)_{\rm R}$ symmetry. 
LRSM provides a natural framework for light neutrino mass generation, which includes rich physical insights. 
The simultaneous presence of right-handed neutrino and scalar triplet enables type I and/or type II seesaw for neutrino mass generation. 
Consequently, both type-I and type-II seesaw contributions to the effective light neutrino mass matrix $M_{\nu}$ are present:
\begin{equation}
\label{eq:2}
M_{\nu} = M_{\nu}^{\rm II} + M_{\nu}^{\rm I} = rM_{\rm R} - M_{\rm D}M_{\rm R}^{-1}M_{\rm D}^{T},
\end{equation}
where $r = v_{\rm L}/v_{\rm R}$ is a scaling factor, with $v_{\rm L}$ and $v_{\rm R}$ denoting the left scalar triplet and right scalar triplet VEVs, respectively.
In the mass basis of charged leptons, $M_\nu$ can be diagonalized by the PMNS matrix $U$ as follows~\cite{41,42,43}:
\begin{align}
\label{eq:3}
U = \begin{pmatrix}
c_{12}c_{13} & s_{12}c_{13} & s_{13}e^{-i \delta} \\
-s_{12}c_{23} -c_{12}s_{13}s_{23} e^{i \delta} & c_{12}c_{23} - s_{12}s_{13}s_{23} e^{i \delta} & c_{13}s_{23} \\
s_{12}s_{23} -c_{12}s_{13}c_{23} e^{i \delta} & -c_{12}s_{23} - s_{12}s_{13}c_{23} e^{i \delta} & c_{13}c_{23} 
\end{pmatrix}
\cdot\begin{pmatrix}
e^{i\rho} & 0 & 0 \\
0 & e^{i\sigma} & 0 \\
0 & 0 & 1
\end{pmatrix},
\end{align}
where $c_{ij} = \cos \theta_{ij}$ and $s_{ij} = \sin  \theta_{ij}$ (for~$ij$ = 12, 13, 23). 
Here, $\delta$ denotes the Dirac CP phase, while $\rho$ and $\sigma$ represent two Majorana phases.
Since neutrino oscillation experiments are insensitive to the Majorana phases, they are completely unknown.
The global analysis of neutrino oscillation data is available in Refs.~\cite{44,45,46}. To ensure consistency, the numerical results from Ref.~\cite{44} (as summarized in Table~\ref{tab:table1}~) are adopted serving as a baseline for the following numerical analysis.
Considering the significant uncertainty associated with $\delta$, we treat it as an independent free parameter in our subsequent analysis.
On the other hand, the LRSM provides two possible mechanisms for generating the lepton asymmetry, depending on the relative masses of the lightest heavy neutrino and the scalar triplet.
There exist some successful implementations of leptogenesis within LRSM, inspired by SO(10) grand unified theory (GUT)~\cite{26,29}, employing various strategies to constrain the free parameters.
The various leptogenesis scenarios in LRSM are discussed in Ref.~\cite{45}.
\begin{table}[h]
\centering
\caption{\label{tab:table1} The best-fit points (bfp) and 3$\sigma$ ranges of five neutrino oscillation parameters extracted from a global analysis of the existing
neutrino oscillation data as of NuFit-6.0~\cite{44}, where $\Delta m^2_{3\ell} = \Delta m^2_{31}>0$ for the NO case and $\Delta m^2_{3\ell} = \Delta m^2_{32}<0$ for the IO case.}
\begin{footnotesize}
\begin{tabular}{|c|cc|cc|}
\hline
\multirow{2}{*}{\begin{tabular}[c]{@{}c@{}} Neutrino oscillation\\ parameters \end{tabular}} 
& \multicolumn{2}{c|}{Normal Ordering} & \multicolumn{2}{c|}{Inverted Ordering } \\
\cline{2-5}
& bfp  & $3\sigma$ range & bfp  & $3\sigma$ range  \\
\cline{1-5}
\rule{0pt}{4mm}\ignorespaces
$\sin^2\theta^{}_{12}$ & 0.307 & $0.275 \to 0.345$ & 0.308 & $0.275 \to 0.345$  \\[1mm]
$\sin^2\theta^{}_{23}$ & 0.561 & $0.430 \to 0.596$ & 0.562 & $0.437 \to 0.597$  \\[1mm]
$\sin^2\theta^{}_{13}$ & 0.02195 & $0.02023 \to 0.02376$ & 0.02224 & $0.02053 \to 0.02397$  \\[1mm]
$\Delta m^2_{21}/(10^{-5}~{\rm eV}^2)$ & 7.49 & $6.92 \to 8.05$ & 7.49 & $6.92 \to 8.05$  \\[1mm]
$\Delta m^2_{3\ell}/(10^{-3}~{\rm eV}^2)$ & 2.534 & $2.463 \to 2.606$ & -2.510 & $-2.584 \to -2.438$  \\[1mm]
\hline
\end{tabular}
\end{footnotesize}
\end{table}

Based on the above experimental results and the existing theoretical framework, 
this work is devoted to studying, within the LRSM framework and in the basis where the Dirac neutrino mass matrix $M_{\rm D}$ is diagonal, 
which patterns of the three-zero textures of the Majorana neutrino mass matrix $M_{\rm R}$ are consistent with the latest neutrino oscillation data, and to investigating their implications for leptogenesis. The motivation of this work can be summarized as follows: 
texture of the light neutrino mass matrix $M_{\nu}$ with more than two zeros have already been excluded by current neutrino oscillation data.
Given that $M_{\nu}$ is derived from $M_{\rm R}$, we consider the Dirac mass matrix $M_{\rm D}$ with three diagonal entries treated as model-independent parameters, and include both type-I and type-II seesaw contributions within the LRSM framework.
In this setup, are there any patterns of the light neutrino mass matrix $M_{\nu}$ that are consistent with current neutrino oscillation data?
Furthermore, we investigate whether the selected viable patterns can successfully realize leptogenesis.

The remaining parts of this paper are organized as follows.  
In section~\ref{section:2}~we investigate the viable light neutrino mass matrix $M_{\nu}$ patterns that are consistent with current neutrino oscillation data, which arise from a combination of type-I and type-II seesaw mechanisms. 
Based on these viable patterns, the relationships among the heavy neutrino mass spectrum, the Yukawa couplings, and the scale factor $r$ are presented.
In section~\ref{section:3}~we study leptogenesis induced by the decays of heavy neutrinos for these viable patterns.
Finally, the summary of our main results will be given in section~\ref{section:4}.

\section{Three-zero textures of $M_{\rm R}$}
\label{section:2}
Since $M_{\rm R}$ is a complex symmetric matrix, it contains of six independent components. 
Consequently, there are $C_{6}^{3}$ = 20 possible configurations for $M_{\rm R}$ with three-zero textures, which are listed in Table~\ref{tab:table2}.
From Table~\ref{tab:table2}, we find that the ranks of the matrices $M_{\rm R8}$, $M_{\rm R9}$, $M_{\rm R10}$, $M_{\rm R17}$, $M_{\rm R18}$ and $M_{\rm R19}$ are all less than 3. 
These cases are excluded because they fail to satisfy the type-I seesaw formula.
By substituting the remaining 14 $M_{\rm R}$ matrices (which have non-zero determinants) into Eq.~\eqref{eq:2}, 
we obtain the effective light neutrino mass matrices presented in Table~\ref{tab:table3}.
\begin{table}[h]
\centering
\caption{\label{tab:table2} 20 three-zero textures of Majorana neutrino mass matrix $M_{\rm R}$.}
\footnotesize 
\renewcommand{\arraystretch}{1.0} 
$\begin{array}{|c|c|c|c|}
\hline
M_{\rm R1} = \begin{pmatrix} 0 & a & b \\ a & 0 & c \\ b & c & 0 \end{pmatrix} &
M_{\rm R2} = \begin{pmatrix} 0 & a & 0 \\ a & 0 & b \\ 0 & b & c \end{pmatrix} &
M_{\rm R3} = \begin{pmatrix} 0 & 0 & a \\ 0 & b & c \\ a & c & 0 \end{pmatrix} &
M_{\rm R4} = \begin{pmatrix} 0 & a & b \\ a & 0 & 0 \\ b & 0 & c \end{pmatrix} \\ 
\hline
M_{\rm R5} = \begin{pmatrix} 0 & a & c \\ a & b & 0 \\ c & 0 & 0 \end{pmatrix} &
M_{\rm R6} = \begin{pmatrix} a & 0 & b \\ 0 & 0 & c \\ b & c & 0 \end{pmatrix} &
M_{\rm R7} = \begin{pmatrix} a & b & 0 \\ b & 0 & c \\ 0 & c & 0 \end{pmatrix} &
M_{\rm R8} = \begin{pmatrix} 0 & 0 & a \\ 0 & 0 & b \\ a & b & c \end{pmatrix} \\ 
\hline
M_{\rm R9} = \begin{pmatrix} 0 & a & 0 \\ a & b & c \\ 0 & c & 0 \end{pmatrix} &
M_{\rm R10} = \begin{pmatrix} a & b & c \\ b & 0 & 0 \\ c & 0 & 0 \end{pmatrix} &
M_{\rm R11} = \begin{pmatrix} a & b & 0 \\ b & 0 & 0 \\ 0 & 0 & c \end{pmatrix} &
M_{\rm R12} = \begin{pmatrix} a & 0 & b \\ 0 & c & 0 \\ b & 0 & 0 \end{pmatrix} \\ 
\hline
M_{\rm R13} = \begin{pmatrix} 0 & a & 0 \\ a & b & 0 \\ 0 & 0 & c \end{pmatrix} &
M_{\rm R14} = \begin{pmatrix} 0 & 0 & a \\ 0 & b & 0 \\ a & 0 & c \end{pmatrix} &
M_{\rm R15} = \begin{pmatrix} a & 0 & 0 \\ 0 & b & c \\ 0 & c & 0 \end{pmatrix} &
M_{\rm R16} = \begin{pmatrix} a & 0 & 0 \\ 0 & 0 & b \\ 0 & b & c \end{pmatrix} \\ 
\hline
M_{\rm R17} = \begin{pmatrix} a & b & 0 \\ b & c & 0 \\ 0 & 0 & 0 \end{pmatrix} &
M_{\rm R18} = \begin{pmatrix} a & 0 & b \\ 0 & 0 & 0 \\ b & 0 & c \end{pmatrix} &
M_{\rm R19} = \begin{pmatrix} 0 & 0 & 0 \\ 0 & a & b \\ 0 & b & c \end{pmatrix} &
M_{\rm R20} = \begin{pmatrix} a & 0 & 0 \\ 0 & b & 0 \\ 0 & 0 & c \end{pmatrix} \\ 
\hline
\end{array}$
\end{table}
\begin{table}[h]
\centering
\caption{\label{tab:table3} Fourteen patterns of light neutrino mass matrices characterized by a non-vanishing det($M_{\rm R}$).}
\begin{tabular}{|Oc|Oc|}
\hline
$M_{\nu1} = \begin{pmatrix}
\frac{c \, y_1^2}{2 \, a \, b} & a \, r - \frac{y_1 \, y_2}{2 \, a} & b \, r - \frac{y_1 \, y_3}{2 \, b} \\
a \, r - \frac{y_1 \, y_2}{2 \, a} & \frac{b \, y_2^2}{2 \, a \, c} & c \, r - \frac{y_2 \, y_3}{2 \, c} \\
b \, r - \frac{y_1 \, y_3}{2 \, b} & c \, r - \frac{y_2 \, y_3}{2 \, c} & \frac{a \, y_3^2}{2 \, b \, c}
\end{pmatrix}$ 
& 
$M_{\nu2} = \begin{pmatrix}
-\frac{b^2 \, y_1^2}{a^2 \, c} & a \, r - \frac{y_1 \, y_2}{a} & \frac{b \, y_1 \, y_3}{a \, c} \\
a \, r - \frac{y_1 \, y_2}{a} & 0 & b \, r \\
\frac{b \, y_1 \, y_3}{a \, c} & b \, r & c \, r - \frac{y_3^2}{c}
\end{pmatrix}$ \\
\hline
$M_{\nu3} = \begin{pmatrix}
-\frac{c^2 \, y_1^2}{a^2 \, b} & \frac{c \, y_1 \, y_2}{a \, b} & a \, r - \frac{y_1 \, y_3}{a} \\
\frac{c \, y_1 \, y_2}{a \, b} & b \, r - \frac{y_2^2}{b} & c \, r \\
a \, r - \frac{y_1 \, y_3}{a} & c \, r & 0
\end{pmatrix}$ 
& 
$M_{\nu4} = \begin{pmatrix}
0 & a \, r - \frac{y_1 \, y_2}{a} & b \, r \\
a \, r - \frac{y_1 \, y_2}{a} & -\frac{b^2 \, y_2^2}{a^2 \, c} & \frac{b \, y_2 \, y_3}{a \, c} \\
b \, r & \frac{b \, y_2 \, y_3}{a \, c} & c \, r - \frac{y_3^2}{c}
\end{pmatrix}$ \\
\hline
$M_{\nu5} = \begin{pmatrix}
0 & a \, r & c \, r - \frac{y_1 \, y_3}{c} \\
a \, r & b \, r - \frac{y_2^2}{b} & \frac{a \, y_2 \, y_3}{b \, c} \\
c \, r - \frac{y_1 \, y_3}{c} & \frac{a \, y_2 \, y_3}{b \, c} & -\frac{a^2 \, y_3^2}{b \, c^2}
\end{pmatrix}$
&
$M_{\nu6} = \begin{pmatrix}
a \, r - \frac{y_1^2}{a} & \frac{b \, y_1 \, y_2}{a \, c} & b \, r \\
\frac{b \, y_1 \, y_2}{a \, c} & -\frac{b^2 \, y_2^2}{a \, c^2} & c \, r - \frac{y_2 \, y_3}{c} \\
b \, r & c \, r - \frac{y_2 \, y_3}{c} & 0
\end{pmatrix}$\\
\hline
$M_{\nu7} = \begin{pmatrix}
a \, r - \frac{y_1^2}{a} & b \, r & \frac{b \, y_1 \, y_3}{a \, c} \\
b \, r & 0 & c \, r - \frac{y_2 \, y_3}{c} \\
\frac{b \, y_1 \, y_3}{a \, c} & c \, r - \frac{y_2 \, y_3}{c} & -\frac{b^2 \, y_3^2}{a \, c^2}
\end{pmatrix}$
&
$M_{\nu11} = \begin{pmatrix}
a \, r & b \, r - \frac{y_1 \, y_2}{b} & 0 \\
b \, r - \frac{y_1 \, y_2}{b} & \frac{a \, y_2^2}{b^2} & 0 \\
0 & 0 & c \, r - \frac{y_3^2}{c}
\end{pmatrix}$\\
\hline
$M_{\nu12} = \begin{pmatrix}
a \, r & 0 & b \, r - \frac{y_1 \, y_3}{b} \\
0 & c \, r - \frac{y_2^2}{c} & 0 \\
b \, r - \frac{y_1 \, y_3}{b} & 0 & \frac{a \, y_3^2}{b^2}
\end{pmatrix}$
&
$M_{\nu13} = \begin{pmatrix}
\frac{b \, y_1^2}{a^2} & a \, r - \frac{y_1 \, y_2}{a} & 0 \\
a \, r - \frac{y_1 \, y_2}{a} & b \, r & 0 \\
0 & 0 & c \, r - \frac{y_3^2}{c}
\end{pmatrix}$\\
\hline
$M_{\nu14} = \begin{pmatrix}
\frac{c \, y_1^2}{a^2} & 0 & a \, r - \frac{y_1 \, y_3}{a} \\
0 & b \, r - \frac{y_2^2}{b} & 0 \\
a \, r - \frac{y_1 \, y_3}{a} & 0 & c \, r
\end{pmatrix}$
&
$M_{\nu15} = \begin{pmatrix}
a \, r - \frac{y_1^2}{a} & 0 & 0 \\
0 & b \, r & c \, r - \frac{y_2 \, y_3}{c} \\
0 & c \, r - \frac{y_2 \, y_3}{c} & \frac{b \, y_3^2}{c^2}
\end{pmatrix}$\\
\hline
$M_{\nu16} = \begin{pmatrix}
a \, r - \frac{y_1^2}{a} & 0 & 0 \\
0 & \frac{c \, y_2^2}{b^2} & b \, r - \frac{y_2 \, y_3}{b} \\
0 & b \, r - \frac{y_2 \, y_3}{b} & c \, r
\end{pmatrix}$
&
$M_{\nu20} = \begin{pmatrix}
a \, r - \frac{y_1^2}{a} & 0 & 0 \\
0 & b \, r - \frac{y_2^2}{b} & 0 \\
0 & 0 & c \, r - \frac{y_3^2}{c}
\end{pmatrix}$\\
\hline
\end{tabular}
\end{table}
The patterns $M_{\nu11}$, $M_{\nu12}$, $M_{\nu13}$, $M_{\nu14}$, $M_{\nu15}$, $M_{\nu16}$, $M_{\nu20}$ are excluded because they are inconsistent with the constraints from current neutrino oscillation data.
This suggests that neutrinos either involve the mixing of only two flavors or no mixing at all.
Based on the above considerations, we now focus on the seven patterns $M_{\rm R1}$, $M_{\rm R2}$, $M_{\rm R3}$, $M_{\rm R4}$, $M_{\rm R5}$, $M_{\rm R6}$ and $M_{\rm R7}$.

In the mass basis of charged leptons, considered in this work, we can also write the light neutrino mass matrix as
\begin{equation}
\label{eq:4}
M_{\nu} = U M_{\nu}^{\rm diag} U^{T},
\end{equation}
where $M_{\nu}^{\rm diag}$ = diag($m_{1}$, $m_{2}$, $m_{3}$). Here, $m_{1}$, $m_{2}$ and $m_{3}$ are three light neutrino masses eigenvalues. 
Note that there are two possible neutrino mass orderings: the NO case with $m_{1}$ $<$ $m_{2}$ $<$ $m_{3}$,
and the IO case with $m_{3}$ $<$ $m_{1}$ $<$ $m_{2}$. 
The neutrino mass squared differences represents $\Delta m_{ij}^{2} \equiv m_{i}^{2}-m_{j}^{2}$, whose latest global fit values are shown in Table~\ref{tab:table1}.
From Table~\ref{tab:table3}, we note that all matrix elements of $M_{\nu1}$ are non-zero, whereas the matrices $M_{\nu2}$, $M_{\nu3}$, $M_{\nu4}$, $M_{\nu5}$, $M_{\nu6}$ and $M_{\nu7}$ each contain one zero element. 
These zero elements can be used to constrain certain low-energy neutrino parameters, such as $\rho$ and $\sigma$.
Consequently, we discuss two separate cases in the following subsections: $M_{\nu1}$ and $M_{\nu,i}$~($i = 2,\cdot\cdot\cdot,7$).

\subsection{$M_{\nu1}$ case}
\begin{figure}[h]
\centering
  \includegraphics[width=.42\textwidth,height=0.21\textheight]{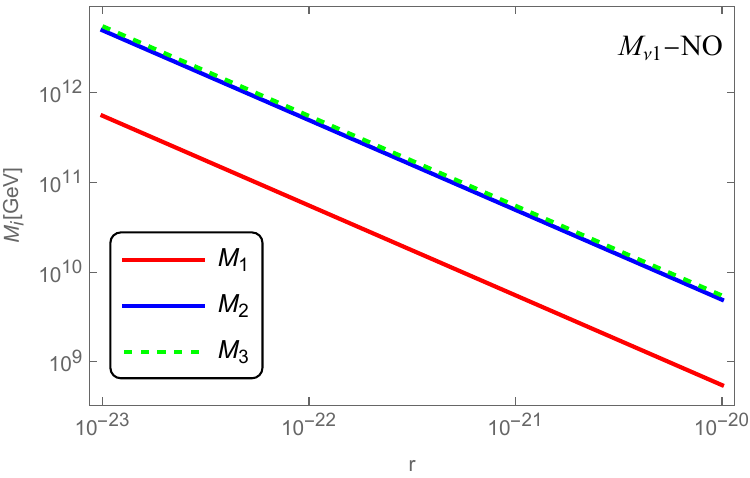}
  \hfill
  \includegraphics[width=.42\textwidth,height=0.21\textheight]{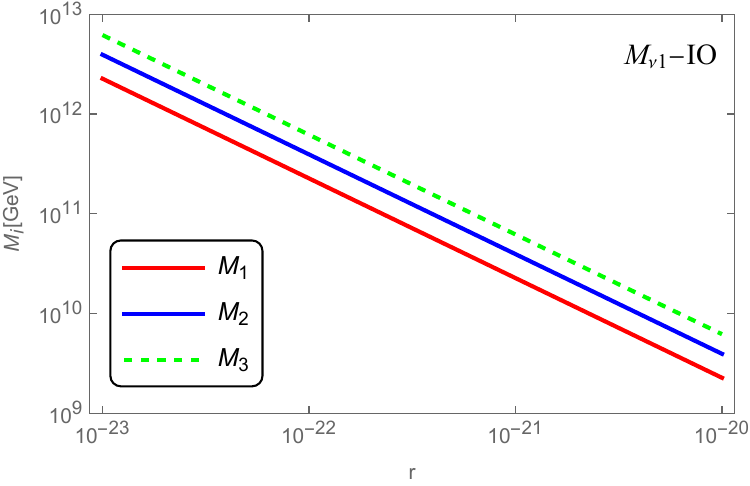}
\caption{\label{fig:1} For $M_{\nu1}$ case, the dependence on $r$ of the heavy right-handed neutrino mass spectrum.}
\end{figure}

\begin{figure}[h]
\centering
  \includegraphics[width=.42\textwidth,height=0.21\textheight]{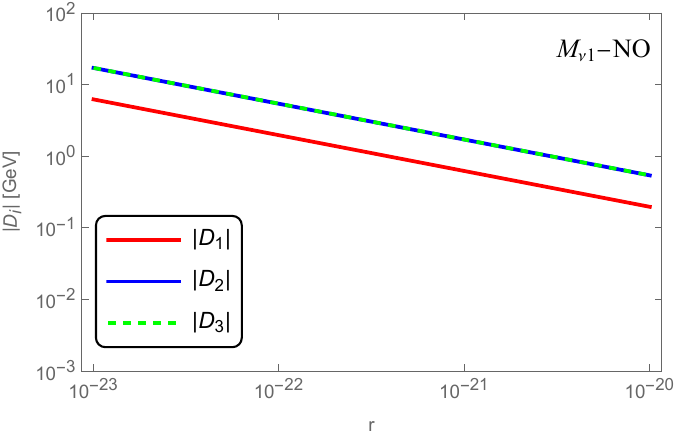}
  \hfill
  \includegraphics[width=.42\textwidth,height=0.21\textheight]{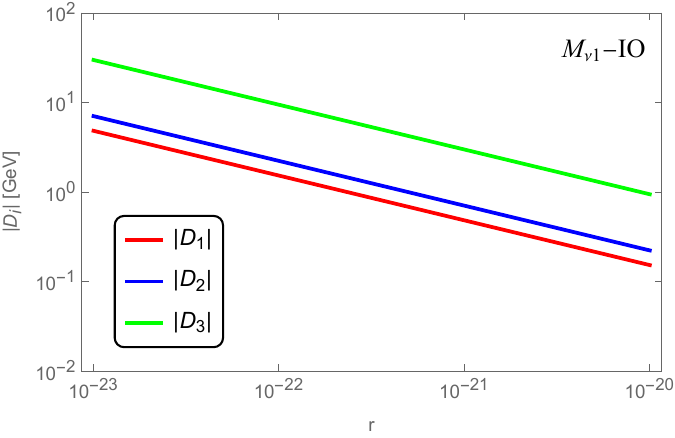}
\caption{\label{fig:2} For $M_{\nu1}$ case, the relationship between the Dirac matrix element $|D_{i}|$ and $r$.}
\end{figure}

First, we examine the case of $M_{\nu1}$. Combining Eq.~\eqref{eq:2} and Eq.~\eqref{eq:4}, we obtain:
\begin{equation}
\label{eq:5}
U M_{\nu}^{diag} U^{T} = 
r \begin{pmatrix} 0 & a & b \\ a & 0 & c \\ b & c & 0 \end{pmatrix} -
\begin{pmatrix}
D_1 & 0 & 0 \\
0 & D_2 & 0 \\
0 & 0 & D_3
\end{pmatrix}
\begin{pmatrix} 0 & a & b \\ a & 0 & c \\ b & c & 0 \end{pmatrix}^{-1}
\begin{pmatrix}
D_1 & 0 & 0 \\
0 & D_2 & 0 \\
0 & 0 & D_3
\end{pmatrix}.
\end{equation}
At this stage, the lightest active neutrino mass (corresponding to $m_{1}$ for NO case and $m_{3}$ for IO case), 
along with the Dirac CP phase $\delta$ and two Majorana phases $\rho$ and $\sigma$, are varied randomly within the ranges [0.001, 0.1] eV, $[0,~2\pi]$, 
$[0,~\pi]$, respectively.
The neutrino oscillation parameters $\theta_{12}$, $\theta_{13}$, $\theta_{23}$, $\Delta m^2_{21}$ and $\Delta m^2_{3\ell}$ are taken as their best-fit values as listed in Table~\ref{tab:table1}.
Matching the corresponding matrix entries in two sides of Eq.~\eqref{eq:5} yields a series of $r$-dependent functions: $a(r)$, $b(r)$, $c(r)$, $D_{1}(r)$, $D_{2}(r)$, and $D_{3}(r)$.
By performing the Autonne-Takagi diagonalization on the complex symmetric matrix $M_{\rm R1}(r)$, 
we obtain the right-handed neutrino masses $M_{1}(r)$, $M_{2}(r)$, $M_{3}(r)$ for both NO and IO cases, as shown in Figure~\ref{fig:1}. 
Simultaneously, the relationships among the Dirac matrix elements $|D_{1}(r)|$, $|D_{2}(r)|$ and $|D_{3}(r)|$ are presented in Figure~\ref{fig:2} for both cases.
Figure~\ref{fig:1} exhibits a clear inverse power-law relationship: as $r$ increases from $10^{-23}$ to $10^{-20}$, the three mass eigenvalues decrease linearly on a log-log scale. In both cases, $M_{1}$ is significantly lighter.
The Yukawa coupling strengths shown in Figure~\ref{fig:2} also decrease when $r$ increases. In the NO case, $|D_{2}(r)|$ and $|D_{3}(r)|$ are relatively close to each other, 
with $|D_{1}(r)|$ being the smallest one. 
However, in the IO case, a more pronounced hierarchical structure is observed, where $|D_{3}(r)|$ is significantly larger than $|D_{1}(r)|$ and $|D_{2}(r)|$.
\subsection{$M_{\nu,i}$~($i = 2,\cdot\cdot\cdot,7$)~case}
\begin{figure}[h]
\centering
\includegraphics[width=0.95\textwidth,height=0.32\textheight]{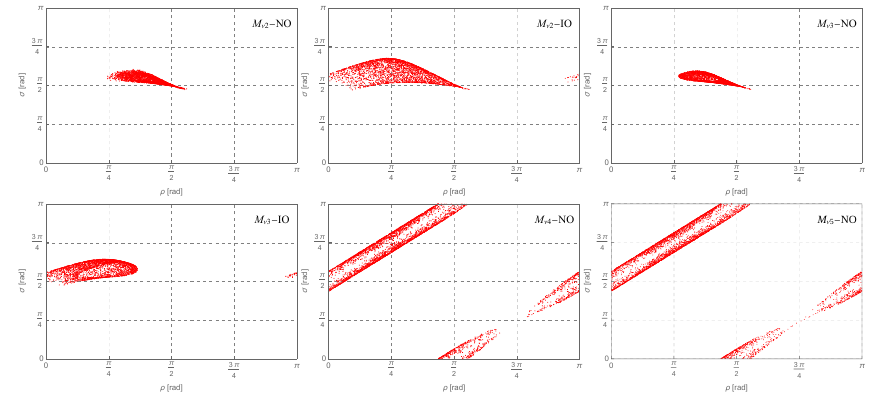}
\caption{\label{fig:3} For $M_{\nu2}$, $M_{\nu3}$, $M_{\nu4}$ and $M_{\nu5}$ cases, the relationship between the Majorana phases $\rho$ and $\sigma$.}
\end{figure}

Next, we investigate the cases for $M_{\nu,i}~(i = 2,\cdot\cdot\cdot,7)$. The procedure is outlined as follows:
(1)~The lightest active neutrino mass ($m_{1}$ for NO case and $m_{3}$ for IO case) and the Dirac CP phase $\delta$ are randomly sampled within the ranges [0.001, 0.1] eV and [0, 2$\pi$], respectively. 
(2)~The parameters $\theta_{12}$, $\theta_{13}$, $\theta_{23}$, $\Delta m^2_{21}$ and $\Delta m^2_{3\ell}$ are taken as their best-fit values or 3$\sigma$ ranges as listed in Table~\ref{tab:table1}.
Substituting these values into Eq.~\eqref{eq:4}, we obtain the correlation between the Majorana phases $\rho$ and $\sigma$, as shown in Figure~\ref{fig:3}.
Figure~\ref{fig:3} indicates that the allowed regions for these phases are highly constrained and strongly model-dependent.
For the $M_{\nu2}$ and $M_{\nu3}$ patterns, the allowed parameter space appears as localized clusters of scatter points. 
In the NO case, the points are more concentrated, whereas in the IO case, the allowed regions are broader, with $\sigma$ showing a larger spread than $\rho$.
By contrast, the patterns of $M_{\nu4}$ and $M_{\nu5}$ in the NO case exhibit a pronounced linear correlation between $\rho$ and $\sigma$.
The data points are distributed along narrow bands, indicating a strong correlation between the two Majorana phases.

\begin{figure}[h]
\centering
\includegraphics[width=0.97\textwidth,height=0.32\textheight]{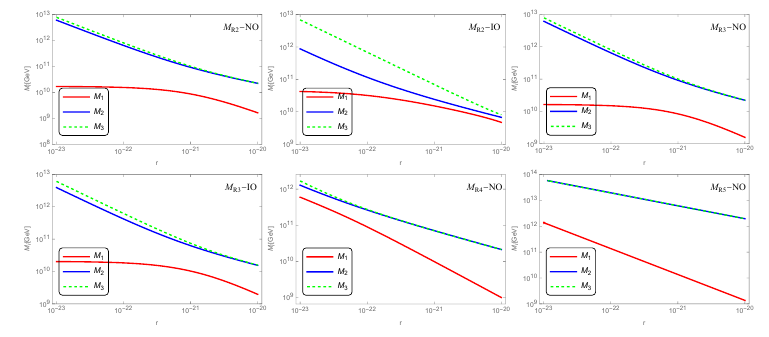}
\caption{\label{fig:4} For $M_{\nu2}$, $M_{\nu3}$, $M_{\nu4}$ and $M_{\nu5}$ cases, the dependence on $r$ of the heavy right-handed neutrino mass spectrum.}
\end{figure}

\begin{figure}[h]
\centering
\includegraphics[width=0.97\textwidth,height=0.38\textheight]{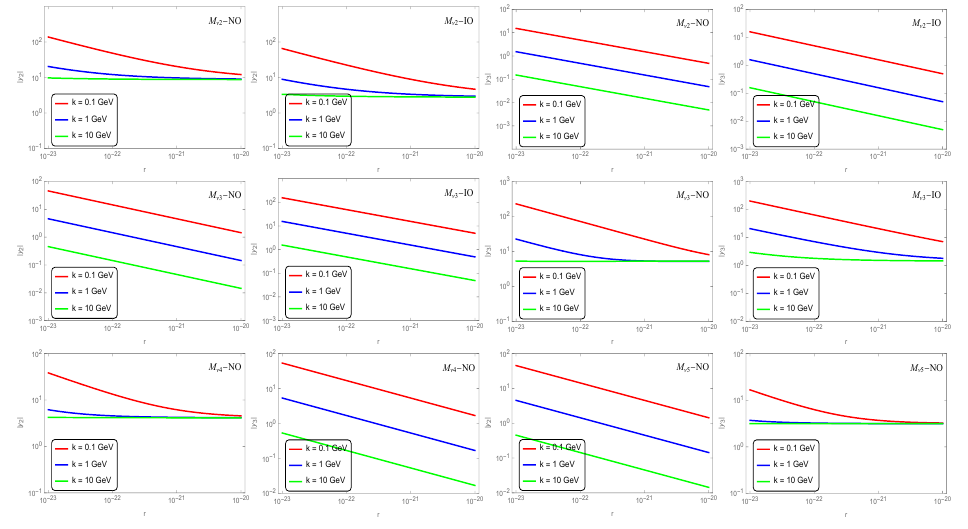}
\caption{\label{fig:5} The dependence of $r$ on the Yukawa coupling $|y_{2}|$ and $|y_{3}|$ in $M_{\nu2}$, $M_{\nu3}$, $M_{\nu4}$ and $M_{\nu5}$ cases.}
\end{figure}

Specifically, we take $M_{\nu2}$ as an illustrative example. 
By combining Eq.~\eqref{eq:2} and Eq.~\eqref{eq:4} and adopting the following parameterization, we obtain
\begin{equation}
\label{eq:6}
U M_{\nu}^{diag} U^{T} = 
r \begin{pmatrix} 0 & a & 0 \\ a & 0 & b \\ 0 & b & c \end{pmatrix} -
k^{2} \begin{pmatrix}
1 & 0 & 0 \\
0 & y_2 & 0 \\
0 & 0 & y_3
\end{pmatrix}
\begin{pmatrix} 0 & a & 0 \\ a & 0 & b \\ 0 & b & c \end{pmatrix}^{-1}
\begin{pmatrix}
1 & 0 & 0 \\
0 & y_2 & 0 \\
0 & 0 & y_3
\end{pmatrix}.
\end{equation}
Here, $k$ denotes the overall scale of $M_{\rm D}$, while $y_{2}$ and $y_{3}$ are dimensionless parameters. 
We assign several benchmark values to $k$, namely 0.1 GeV, 1 GeV, and 10 GeV. 
By matching the corresponding non-zero matrix elements on both sides of Eq.~\eqref{eq:6}, 
we derive a set of analytical expressions for the parameters as functions of $r$, namely $a(r)$, $b(r)$, $c(r)$, $y_{2}(r)$, and $y_{3}(r)$. 
Subsequently, by performing the Autonne--Takagi diagonalization of the complex symmetric matrix $M_{\rm R,i}(r)$,
we obtain the right-handed neutrino masses $M_{1}(r)$, $M_{2}(r)$, $M_{3}(r)$, as shown in Figure~\ref{fig:4}. 
Furthermore, $|y_{2}(r)|$ and $|y_{3}(r)|$ are presented in Figure~\ref{fig:5}.
From Figure~\ref{fig:4}, a common feature across all panels is that the three mass eigenvalues decrease monotonically as the parameter $r$ increases from $10^{-23}$ to $10^{-20}$.
In most scenarios, such as $M_{\rm R2}$ and $M_{\rm R3}$, a clear hierarchical pattern is observed, namely $M_{3}$ $\gg$ $M_{2}$ $\gg$ $M_{1}$.
Notably, in the NO case the mass eigenvalues $M_{2}$ (blue solid line) and $M_{3}$  (green dashed line) are nearly degenerate over the entire range of $r$, while $M_{1}$ remains significantly lighter.
In the IO cases, the mass splitting between $M_{2}$ and $M_{3}$ is more pronounced at smaller values of $r$. 
As $r$ increases, the gap between $M_{2}$ and $M_{3}$ gradually narrows, indicating a tendency toward convergence at larger $r$.

Figure~\ref{fig:5} shows that $|y_{2}|$ and $|y_{3}|$ generally decrease with increasing $r$, 
indicating that the relative coupling strengths in the Dirac sector are highly sensitive to the scale of $r$.
In addition, a clear stratification with respect to the values of $k$ can be observed. 
For a fixed $r$, a smaller $k$ value (0.1 GeV, red line) leads to significantly larger ratios than a larger value of $k$ (10 GeV, green line). 
This suggests an inverse correlation between the energy scale $k$ and the magnitude of these matrix element. 
In particular, for $|y_{2}|$ in the $M_{\nu2}$ and $M_{\nu4}$ patterns in the NO case, 
the curves corresponding to different values of $k$ tend to converge as $r$ approaches $10^{-20}$. 
This behavior implies that the influence of the parameter $k$ on the Dirac matrix becomes less dominant in the large-$r$ regime.

In view of the above analysis, we summarize the $M_{\nu}$ patterns that are consistent with the global-fit results of NuFit 6.0 in Table~\ref{tab:table4}.
These results serve as the basis for exploring leptogenesis originating from the decays of heavy neutrinos.
\begin{table}[htbp]
\centering
\caption{\label{tab:table4} The $M_{\nu}$ patterns that satisfy the global fit values of NuFit 6.0. The bfp and 3$\sigma$ in parentheses represent the best-fit point and the 3$\sigma$ ranges in Table~\ref{tab:table1}, respectively.}
\begin{tabular}{|c|c|c|c|c|c|c|c|}
\hline
\diagbox{mass ordering}{$M_\nu$} & $M_{\nu1}$ & $M_{\nu2}$ & $M_{\nu3}$ & $M_{\nu4}$ & $M_{\nu5}$ & $M_{\nu6}$ & $M_{\nu7}$ \\ \hline
NO & $\checkmark$(bfp) & $\checkmark(3\sigma)$ & $\checkmark$(bfp) & $\checkmark$(bfp) & $\checkmark$(bfp) & $\times$ & $\times$ \\ \hline
IO & $\checkmark$(bfp) & $\checkmark$(bfp)     & $\checkmark$(bfp) & $\times$ & $\times$ & $\times$ & $\times$ \\ \hline
\end{tabular}
\end{table}

\section{Leptogenesis from heavy neutrino decays}
\label{section:3}
In the LRSM, the lepton asymmetry is generated by either the lightest heavy neutrino or the scalar triplet, which depends on their relative masses. 
In this paper, we take the mass of scalar triplet $m_{\Delta} = 4 \times 10^{13}$~GeV as a benchmark value. 
It can be observed from Figure~\ref{fig:1} and Figure~\ref{fig:4} that the right-handed neutrino masses are hierarchical and the lightest one $M_{1}<10^{12}$~GeV.
Therefore, flavor effects must be taken into account in leptogenesis~\cite{47,48}.

\subsection{The CP asymmetries}
\begin{figure}[h]
\centering
\includegraphics[width=0.25\textwidth]{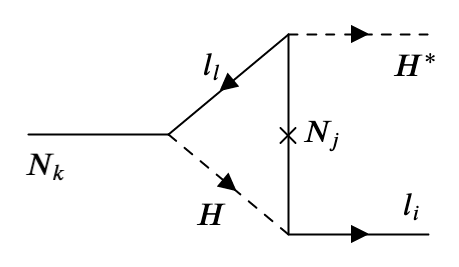}\hspace{1cm}
\includegraphics[width=0.25\textwidth]{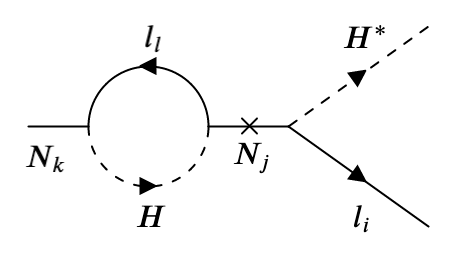}\hspace{1cm}
\includegraphics[width=0.25\textwidth]{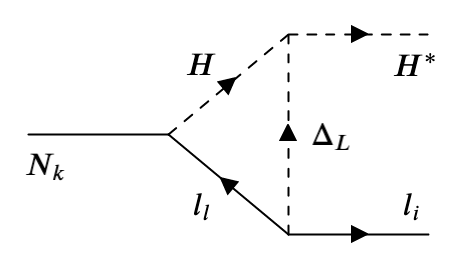}
\caption{\label{fig:6} One loop diagrams for the heavy neutrino decay\cite{38}.}
\end{figure}

The CP asymmetry generated by the decay of the $k$-th heavy Majorana neutrino $N_{k}~(k = 1,2,3)$ is defined as
\begin{equation}
\label{eq:7}
\epsilon_{N_k}=\displaystyle\sum_i \frac{\Gamma(N_k\rightarrow l_i H^*)- \Gamma(N_k\rightarrow \bar{l}_i H)}
{\Gamma(N_k\rightarrow l_i H^*)+ \Gamma(N_k\rightarrow \bar{l}_i H)},
\end{equation}
which is generated by the interference between the tree-level decay amplitude with one-loop amplitudes as shown in Figure.~\ref{fig:6}. 
Here, $l_{i}$~($\bar{l}_{i}$) denotes the lepton~(anti-lepton) doublet, $H$ represents the Higgs doublet, and
$H^{\ast}$ represents hermitian conjugate of the Higgs doublet.

The CP asymmetry generated by interference with the first two one-loop amplitudes is the usual type-I seesaw asymmetry\cite{45}
\begin{equation}
\label{eq:8}
\epsilon_{N_k} =\displaystyle\frac{1}{8\pi}\sum_j \frac{\mathrm{Im} (Y_NY_N^\dagger)_{kj}^2}{\sum_i |(Y_N)_{ki}|^2} \sqrt{x_j} 
\left[ 1-(1+x_j) \mathrm{ln}(1+\frac{1}{x_j})+\frac{1}{1-x_j} \right],
\end{equation}
where $Y_N$ = $\widetilde{M_{\rm D}}$/$v$ (with $v$ = 174 GeV),
and the redefined Dirac mass matrix $\widetilde{M_{\rm D}}$ = $M_{\rm D}$$\tilde{U}$, with $\tilde{U}^{\rm T}M_{\rm R}\tilde{U}$ = diag($M_{1},M_{2},M_{3}$).
Here, $x_j = m_{N_j}^2/m_{N_k}^2$, $m_{N_k} = M_k~(k = 1,2,3)$.
The contribution of the third diagram is~\cite{45}:
\begin{equation}
\label{eq:9}
\epsilon_{N_k}^\Delta =\displaystyle -\frac{1}{2 \pi} 
\frac{\sum_{il} \mathrm{Im}\left[(Y_N)_{ki} (Y_N)_{kl} (Y_\mathrm{T}^*)_{il} \mu\right]}{\sum_i |(Y_N)_{ki}|^2 m_{N_k}} 
\left[ 1- \frac{m_\Delta^2}{m_{N_k}^2} \mathrm{ln}\left(1+\frac{m_{N_k}^2}{m_\Delta^2}\right)  \right],
\end{equation}
where $Y_{T} = M_{\rm R} / v_{\rm R}$, and $\mu = v_{\rm L}m_{\Delta}^{2}/v^{2}$.
Thus, the expression for the total CP asymmetry is given by:
\begin{equation}
\label{eq:10}
\epsilon_{k} \equiv \epsilon_{N_k} + \epsilon_{N_k}^\Delta.
\end{equation}

\subsection{Some basic formulas of leptogenesis}
For hierarchical right-handed neutrino masses, the final baryon asymmetry originates primarily from the decays of $N^{}_1$, Any primordial asymmetry generated by $N_{2}$ and $N_{3}$ is typically washed out by $N^{}_1$ interactions before it can be converted. In the unflavored regime the final baryon asymmetry from $N_{1}$ is given by\cite{22,23,24,25}
\begin{equation}
\label{eq:11}
Y^{}_{\rm B} = c_{1} r_{1} \epsilon_{1} \kappa(\widetilde m_{1}),
\end{equation}
where $c_{1} = -~28/79$ denotes the conversion efficiency of the lepton-to-baryon asymmetry through sphaleron transitions within the SM framework.
The factor $r_{1} \simeq 4 \times 10^{-3}$ represents the ratio of the equilibrium number density of $N_{1}$ to the entropy density.
Furthermore, the total CP asymmetry $\epsilon_{1}$, as defined in Eq.~\eqref{eq:10}, 
results from the summation over the individual flavored CP asymmetries $\epsilon_{1\alpha}$~(where $\alpha = e, \mu, \tau$).
\begin{equation}
\label{eq:12}
\begin{aligned}
\epsilon_{1\alpha} = & \frac{1}{8\pi} \sum_{j \neq 1} \frac{\text{Im} \left[ (Y_N)_{\alpha 1} (Y_N Y_N^\dagger)_{1j} (Y_N^\dagger)_{j \alpha} \right]}{\sum_l |(Y_N)_{1l}|^2} \sqrt{x_j} \left[ 1 - (1 + x_j) \ln \left( 1 + \frac{1}{x_j} \right) + \frac{1}{1 - x_j} \right] \\
& - \frac{1}{2\pi} \frac{\sum_l \text{Im} \left[ (Y_N)_{1\alpha} (Y_N)_{1l} (Y_T^* )_{\alpha l} \mu \right]}{\sum_i |(Y_N)_{1i}|^2 m_{N_1}} \left[ 1 - \frac{m_\Delta^2}{m_{N_1}^2} \ln \left( 1 + \frac{m_{N_1}^2}{m_\Delta^2} \right) \right].
\end{aligned}
\end{equation}
Lastly, the efficiency factor $\kappa(\widetilde m_{1}) \leq 1$ characterizes the survival probability of the lepton asymmetry generated via $N_{1}$ decays.
It accounts for depletion caused by both $N_{1}$ inverse decays and various lepton number violating  scattering processes.
Its concrete value is determined by the washout mass parameter
\begin{equation}
\label{eq:13}
\widetilde m^{}_1 = \sum^{}_\alpha \widetilde m^{}_{1 \alpha} = \sum^{}_\alpha  \frac{|(M^{}_{\rm D})^{}_{\alpha 1}|^2}{M^{}_1},
\end{equation}
which is typically determined by solving the relevant Boltzmann equations.
Within the strong washout regime—representative of most realistic leptogenesis scenarios—the efficiency factor scales approximately as the inverse of $\widetilde m^{}_1$.
In the 2-flavor regime relevant to our study, the final baryon asymmetry from $N_{1}$ is given by
\begin{equation}
\label{eq:14}
Y^{}_{\rm B}
=  c_{1} r_{1} \left[ \epsilon^{}_{1 \gamma} \kappa \left(\frac{417}{589} \widetilde m^{}_{1 \gamma} \right) + \epsilon^{}_{1 \tau} \kappa \left(\frac{390}{589} \widetilde m^{}_{1 \tau} \right) \right],
\end{equation}
with $\epsilon^{}_{1 \gamma} = \epsilon^{}_{1 e} + \epsilon^{}_{1 \mu}$ and $\widetilde m^{}_{1 \gamma} = \widetilde m^{}_{1 e} + \widetilde m^{}_{1 \mu}$. 

\subsection{Numerical results}
\begin{figure}[h]
\centering
\includegraphics[width=.90\textwidth,height=0.41\textheight]{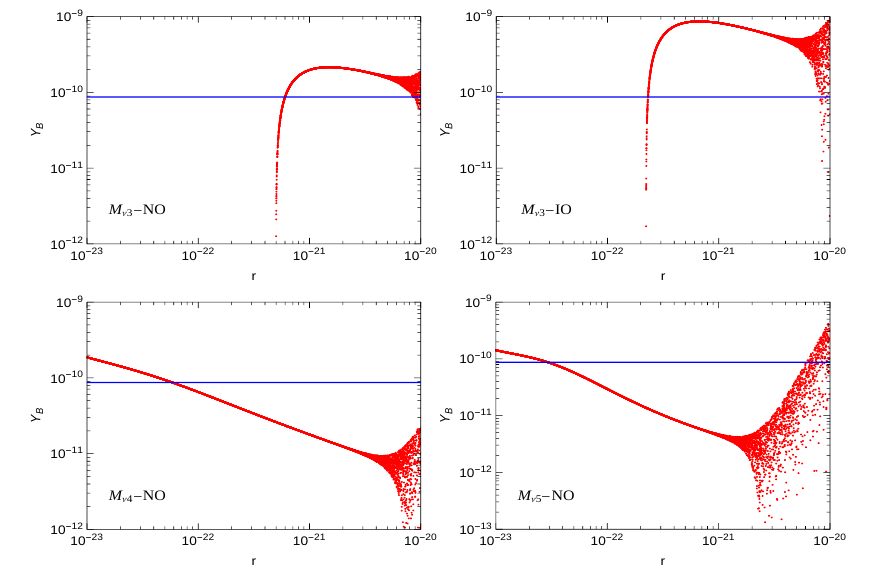}
\caption{\label{fig:7} For $M_{\nu3}$, $M_{\nu4}$ and $M_{\nu5}$ cases, the relationship between the final baryon asymmetry $Y_{\rm B}$ and $r$. The blue horizontal line indicates the observed value $Y_{\rm B} \approx 8.7 \times 10^{-11}$.}
\end{figure}

\begin{figure}[h]
\centering
\includegraphics[width=.92\textwidth,height=0.28\textheight]{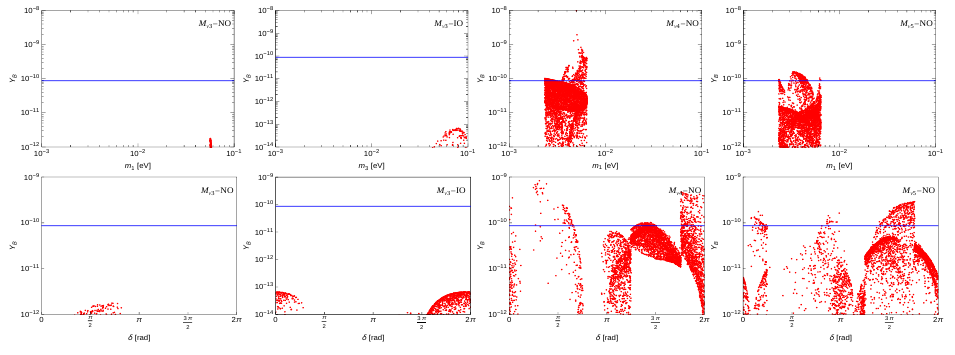}
\caption{\label{fig:8} For $M_{\nu3}$, $M_{\nu4}$ and $M_{\nu5}$ cases, the upper panel shows the relationship between the final baryon asymmetry $Y_{\rm B}$ and the lightest active neutrino mass $m_{1}$  (NO case) or $m_{3}$ (IO case). The lower panel shows the relationship between the final baryon asymmetry $Y_{\rm B}$ and the Dirac CP phase $\delta$.}
\end{figure}

In this subsection, we present the numerical results for the baryon asymmetry $Y_{\rm B}$ for various $M_{\nu}$ configurations by incorporating the light neutrino mass patterns  listed in Table~\ref{tab:table4}, which are consistent with the NuFit 6.0 global-fit data, into Eq.~\eqref{eq:8}~-~Eq.~\eqref{eq:14}.
We find that the $M_{\nu1}$ and $M_{\nu2}$ patterns yield $Y_{\rm B}$ values significantly below the observed value, $Y_{\rm B} \approx 8.68 \times 10^{-11}$~\cite{20}. Therefore, these cases are not shown here. Instead, the results for the $M_{\nu3}$, $M_{\nu4}$, and $M_{\nu5}$ patterns are illustrated in Figure~\ref{fig:7} and Figure~\ref{fig:8}.

Figure~\ref{fig:7} displays the final baryon asymmetry $Y_{\rm B}$ as a function of the parameter $r$ for three distinct light neutrino mass patterns. 
The blue horizontal line indicates the observed value $Y_{\rm B} \approx 8.7 \times 10^{-11}$.
For the $M_{\nu3}$ patterns (upper panels), $Y_{\rm B}$ is negligible at very small $r$ but rises sharply around $r \sim 3 \times 10^{-21}$, quickly exceeding the observed value. 
This behavior indicates that $M_{\nu3}$ patterns require $r$ to lie within a specific range in order to realize successful leptogenesis.
By contrast, for the $M_{\nu4}$ and $M_{\nu5}$ patterns (lower panels), the behavior is converse:
$Y_{\rm B}$ starts above the observed value at $r \sim 10^{-23}$ and generally decreases as $r$ increases. 
As $r$ approaches $\sim 10^{-20}$, the results become increasingly scattered, indicating enhanced sensitivity to other model parameters or CP phases in this regime.

Figure~\ref{fig:8} illustrates the dependence of $Y_{\rm B}$ on the lightest active neutrino mass~($m_{1}$ in the NO case and $m_{3}$ in the IO case) and the Dirac CP phase $\delta$.
In the upper panels, which show the mass dependence, the predicted $Y_{\rm B}$ for the $M_{\nu3}$ pattern lies mostly below the experimental line throughout the scanned mass range.
In contrast, for $M_{\nu4}$ and $M_{\nu5}$ patterns, dense clusters of points reach or exceed the blue line, particularly when the lightest neutrino mass lies in the range $[2 \times 10^{-3}, 6 \times 10^{-3}]$~eV. This indicates that these patterns can more readily accommodate the observed baryon asymmetry.
In the lower panels, which display the dependence on the Dirac CP phase, the relation between $Y_{\rm B}$ and the Dirac CP phase $\delta$ exhibits distinct peak-like structures.
For the $M_{\nu4}$ and $M_{\nu5}$ patterns, successful leptogenesis (corresponding to points above the blue line) is concentrated in specific regions of $\delta$, highlighting the important role of the Dirac CP phase in modulating the CP-violating source responsible for the baryon asymmetry.

\section{Summary}
\label{section:4}
Within the framework of the LRSM, light neutrino masses originate from the combined contribution of Type-I and Type-II seesaw mechanisms. 
Under the assumption that the Dirac neutrino mass matrix $M_{\rm D}$ is diagonal, 
this work systematically investigates 20 types of three-zero textures in the Majorana neutrino mass matrix $M_{\rm R}$. 
Some of these three-zero textures are excluded because they either fail to satisfy the Type-I seesaw mechanism (the rank of $M_{\rm R}$ is less than 3) or are inconsistent with current neutrino oscillation experimental data. 
Our study reveals that only five three-zero textures—namely, $M_{\nu1}$, $M_{\nu2}$, $M_{\nu3}$, $M_{\nu4}$ and $M_{\nu5}$ satisfy the constraints of the latest NuFit 6.0 global fit results.

For the $M_{\nu1}$ case, we examine the relationship between the heavy neutrino mass spectrum $M_{I}$ and the scaling ratio $r$, as well as the correlation between the Yukawa coupling magnitudes $|y_{i}|$ and $r$. 
For the $M_{\nu2}$, $M_{\nu3}$, $M_{\nu4}$, and $M_{\nu5}$ cases, we demonstrate the correlations between Majorana phases $\rho$ and $\sigma$, the heavy neutrino mass spectra versus $r$ and the Yukawa couplings versus $r$. The results indicate strong correlations among model parameters. In particular, the allowed regions for the Majorana CP phases are significantly constrained and depend strongly on the three-zero textures of $M_{\rm R}$.

On this basis, we further investigate leptogenesis originating from the decay of heavy right-handed neutrinos. 
Numerical results show that not all zero textures satisfying the NuFit 6.0 global fit data can successfully explain the baryon asymmetry of the universe.
For instance, the $M_{\nu1}$ and $M_{\nu2}$ patterns fail to produce the observed value of $Y_{\rm B}$. 
The $M_{\nu3}$ pattern can achieve successful leptogenesis within specific $r$ intervals for both NO and IO cases. 
The $M_{\nu4}$ and  $M_{\nu5}$ patterns exhibit viable parameter spaces for successful leptogenesis only in the NO case.

\vspace{0.5cm}

\underline{Acknowledgments} 
\vspace{0.2cm}

This work was supported in part by the National Natural Science Foundation of China under Grant No.~12475112, Liaoning Revitalization Talents Program under Grant No.~XLYC2403152, and the Basic Research Business Fees for Universities in Liaoning Province under Grant No.~LJ212410165050.

\end{document}